# Spin diffusion in geometrically frustrated Heisenberg antiferromagnets

D. L. Huber

Physics Dept., Univ. of Wisconsin-Madison, 1150 University Ave., Madison, WI 53706

## Abstract

We investigate spin diffusion in geometrically frustrated Heisenberg antiferromagnets. It is found that the diffusion constant gradually increases from its high temperature limit as the temperature approaches the peak temperature of the susceptibility. Below the peak, the exponential decrease in the susceptibility leads to a rapid increase in the rate of spin diffusion. Detailed results are presented for the pyrochlore lattice with nearest-neighbor interactions.

# I Introduction

Geometrically frustrated magnets have proven to be an interesting class of magnetic materials [1]. The term 'geometrical frustration' refers to the property that the geometry of the lattice of spins prevents the establishment of long range magnetic order. Most of the theoretical studies of these materials have focussed on the thermodynamic properties. In this paper, we undertake an investigation of the dynamical behavior.

We consider localized spin systems with Heisenberg (isotropic) exchange interactions and focus on the long wavelength – low frequency regime where a hydrodynamic picture is expected to be valid. In the absence of long range order, the hydrodynamic theory for a Heisenberg system leads to diffusive decay of the spin - spin correlation function at long times and large separations. More specifically, in the hydrodynamic regime, the spectral function, which is related to the Fourier transform of the spin – spin correlation function, is predicted to have a Lorentzian form with a width that is proportional to the square of the wave vector. The coefficient of proportionality is the *spin diffusion constant*.

Our focus is on the magnitude and temperature dependence of the diffusion constant. We obtain results for the Heisenberg antiferromagnet with nearest-neighbor interactions on the pyrochlore lattice. General results are obtained for arbitrary spin and numerical results for the case $S = 3/2$. We make use of the fact that the diffusion constant can be expressed as the product of a thermodynamic factor and a parameter that is identified as the decay time of the spin current autocorrelation function. We employ the Generalized Constant Coupling approximation (GCC) [2] to calculate the thermodynamic factor for all $T > 0$, while we use a high temperature approximation to estimate the decay time. The resulting theory is expected to be quantitatively appropriate for temperatures that are greater than the temperature of the peak in the susceptibility and to provide qualitative insight into the behavior of the diffusion constant at lower temperatures. Since much of the general formalism for the analysis of spin diffusion is well established,[3,4,5,6] we will consider mainly the areas that are influenced by the geometrical frustration.

## II Diffusion Constant

The starting point in the analysis is a general expression for the spin diffusion constant that takes the form

$$D = \lim_{k \to 0} k^{-2} \frac{\int_0^\infty dt (dS_z(\mathbf{k},t)/dt, dS_z(-\mathbf{k},0)/dt)}{(S_z(\mathbf{k}), S_z(-\mathbf{k}))}, \tag{1}$$

where $S_z(k)$ is the Fourier transform of the z-component of the spin and the symbol $(A,B)$ denotes the Kubo (susceptibility) inner product

$$(A,B) = \int_0^\beta d\lambda < e^{\lambda H} A e^{-\lambda H} B >_T - \beta < A >_T < B >_T, \tag{2}$$

Here we have $\beta = 1/k_B T$ while the angular brackets denote a thermal average. We have also made the assumption that the $x$, $y$, and $z$ directions are equivalent (macroscopic cubic symmetry) so that the diffusion tensor is proportional to the unit tensor.

Rearranging terms we can rewrite the expression for the diffusion constant as

$$D = \lim_{k \to 0} k^{-2} \overline{\omega}_k^2 \int_0^\infty dt \frac{(dS_z(\mathbf{k},t)dt, dS_z(-\mathbf{k},0))}{(dS_z(\mathbf{k})/dt, dS_z(-\mathbf{k})/dt)}, \tag{3}$$

where the factor in front of the integral is the second moment of the normalized spectral weight function, $F(\mathbf{k},\omega)$, defined by

$$F(\mathbf{k},\omega) = (2\pi)^{-1} \int_{-\infty}^{\infty} dt\, e^{-i\omega t} (S_z(\mathbf{k},t), S_z(-\mathbf{k},0)) / (S_z(\mathbf{k}), S_z(-\mathbf{k})). \qquad (4)$$

That is, we have

$$\bar{\omega}_k^{2n} = \int_{-\infty}^{\infty} \omega^{2n} F(\mathbf{k},\omega) d\omega. \qquad (5)$$

For a Heisenberg magnet, the second moment can be expressed as

$$\bar{\omega}_k^2 = \frac{-2\hbar^{-2}}{(S_z(\mathbf{k}), S_z(-\mathbf{k}))N} \sum_{i,j} J_{ij}(1-\cos(\mathbf{k}\mathbf{r}_{ij})) <S_i^z S_j^z>_T, \qquad (6)$$

where $J_{ij}$ is the exchange interaction between spins separated by the vector $\mathbf{r}_{ij}$ and $N$ is the total number of spins.[6] In a system with nearest-neighbor interactions, Eq. (6) reduces to

$$\bar{\omega}_k^2 = \frac{-2\hbar^{-2} U(T)}{3z\chi(T)N} \sum_{i,j} (\mathbf{k}\mathbf{r}_{ij})^2, \qquad (7)$$

in the long wavelength limit. Here $U(T)$ is the internal energy, $\chi(T)$ is the uniform field susceptibility in units of $g^2\mu_B^2$ ($\chi(T) = (S_z(0), S_z(0))$), and $z$ is the number of nearest neighbors. If we denote the integral in Eq. (3) by $\tau_k$, then the diffusion constant for the pyrochlore lattice can be written as

$$D(T) = \frac{-2U(T)\tau_0(T)\hbar^{-2}l^2}{9\chi(T)}, \qquad (8)$$

where $l$ denotes the separation between nearest neighbors. Equation (8) is the principal result of this section. It expresses the spin diffusion constant as a ratio of thermodynamic quantities and various constants multiplied by a relaxation time.

## III Analysis

In the analysis, we rewrite the diffusion constant as

$$D(T) = A(T)\tau_0(T), \tag{9}$$

where $A(T)$ is defined by the group of parameters multiplying $\tau_0$ in Eq. (8). At high temperatures, $A(T)$ approaches the limiting value

$$A(\infty) = \frac{2}{3} S(S+1) J^2 l^2 \hbar^{-2}. \tag{10}$$

We can use the Generalized Constant Coupling approximation to calculate $A(T)$ at finite temperatures for arbitrary spin. In Fig. 1, we display results for the ratio $A(T)/A(\infty)$ for the case $S = 3/2$ Similar results are obtained for other values of the spin. In all cases, $A(T)$ slowly increases down to temperatures on the order of the peak temperature of the susceptibility which for $S = 3/2$ is $\approx 1.53J$. Below that point, there is a rapid increase due to the exponential decrease in the susceptibility coming from the gap between the ground state and the low-lying excited states. The slow increase at higher temperatures reflects a balance between the increase in the susceptibility and the increase in the magnitude of the internal energy. If we use the high-temperature ($1/T$) approximation for the internal energy and the Curie – Weiss approximation for the susceptibility, we find that $A(T) \approx (1+\Theta_N/T)\, A(\infty)$ for $T \gg SJ$, where $\Theta_N$ denotes the paramagnetic Néel temperature.

The determination of the decay time $\tau_0$ is more complicated. At high temperatures, one can estimate $\tau_0$ by making a Gaussian approximation to the integrand in Eq. (3), *i.e.* exp[$-at^2$], and expressing *a* in terms of the moments of the spectral function. The resulting expression for $\tau_0$ has the form

$$\tau_0(T) = (\pi/2)^{1/2} \lim_{k \to 0} (\overline{\omega}_k^2 / \overline{\omega}_k^4)^{1/2}. \tag{11}$$

At high temperatures (*i.e.*, above the peak in $\chi(T)$), the moments have the form $B_{2n}(T)/T\chi(T)$, where $B_{2n}(T)$ is a slowly varying function of temperature. The *ratio* of the moments is thus also a slowly varying function of temperature. As a consequence, we use the high temperature limit of Eq. (11) to obtain an estimate for $\tau_0$, with the result

$$\tau_0(T) \approx \tau_0(\infty) = (\pi/2)^{1/2} \hbar J^{-1} \left( \frac{6}{32S(S+1)-3} \right)^{1/2}, \tag{12}$$

having used the infinite-temperature moments given in Ref. 6. As expected, $\tau_0(\infty)$ is inversely proportional to the exchange integral. We postpone discussion of these results to the following section.

## IV Discussion

Our results for the diffusion constant are displayed in Eqs. (8) - (10) and (12). Although the detailed results were obtained for the nearest-neighbor Heisenberg antiferromagnet on the pyrochlore lattice, it is expected that the qualitative aspects of the results pertain to other frustrated Heisenberg systems. Taken together, these equations predict a diffusion constant that increases slowly from a finite value at high temperatures. The slow increase continues to temperatures on the order of the peak temperature of the susceptibility. Below the peak, the diffusion constant increases rapidly due to an exponential decrease in the susceptibility. The behavior of the relaxation time $\tau_0$, which is identified with the time integral of the normalized spin current autocorrelation function, may be complicated. At temperatures above the peak, it is expected that $\tau_0$ will be nearly temperature independent. Below the peak, the temperature dependence may become more pronounced. There is growing evidence for the existence of low frequency, non-diffusive modes in frustrated Heisenberg magnets at low temperatures [7,8]. It is possible that the presence of these modes affects the decay of the spin current autocorrelation function by opening up additional relaxation channels thus leading to a decrease in $\tau_0$ as $T \rightarrow 0$.

Our next comment pertains to the extent of the diffusive regime. At high temperatures, the diffusive regime is specified by the condition $kl \ll 1$. However, there is evidence that the diffusive condition becomes more restrictive as $T \rightarrow 0$. The evidence comes from the ratio of the fourth moment to the square of the second moment. If the spectral function has an approximately Lorentzian shape, this ratio will be very large.[6] From the comments about the temperature dependence of the moments made in Sec. III, the ratio will be proportional to $T\chi(T)$, which is a decreasing function of temperature. As a consequence, the diffusive regime is predicted to shrink as the diffusion constant increases.

Concerning the experimental situation, we have found only one reference to spin diffusion in pyrochlore systems.[9] In Ref. 9, studies of the spin dynamics of $Tl_2Mn_2O_7$ showed 'conventional' *i.e.* diffusive behavior at long wavelengths. However, this compound is not representative of frustrated systems in that it undergoes a ferromagnetic transition at 125 K, which is only slightly below the Curie temperature of 170 K.[10] As a consequence, there remains a need for detailed studies of spin diffusion in highly frustrated Heisenberg magnets. A particularly promising candidate for study is $ZnCr_2O_4$.

This system is a spin – 3/2 Heisenberg antiferromagnet on a pyrochlore lattice with dominant nearest – neighbor interactions where the thermodynamic properties are well characterized by the GCC approximation [11][12].

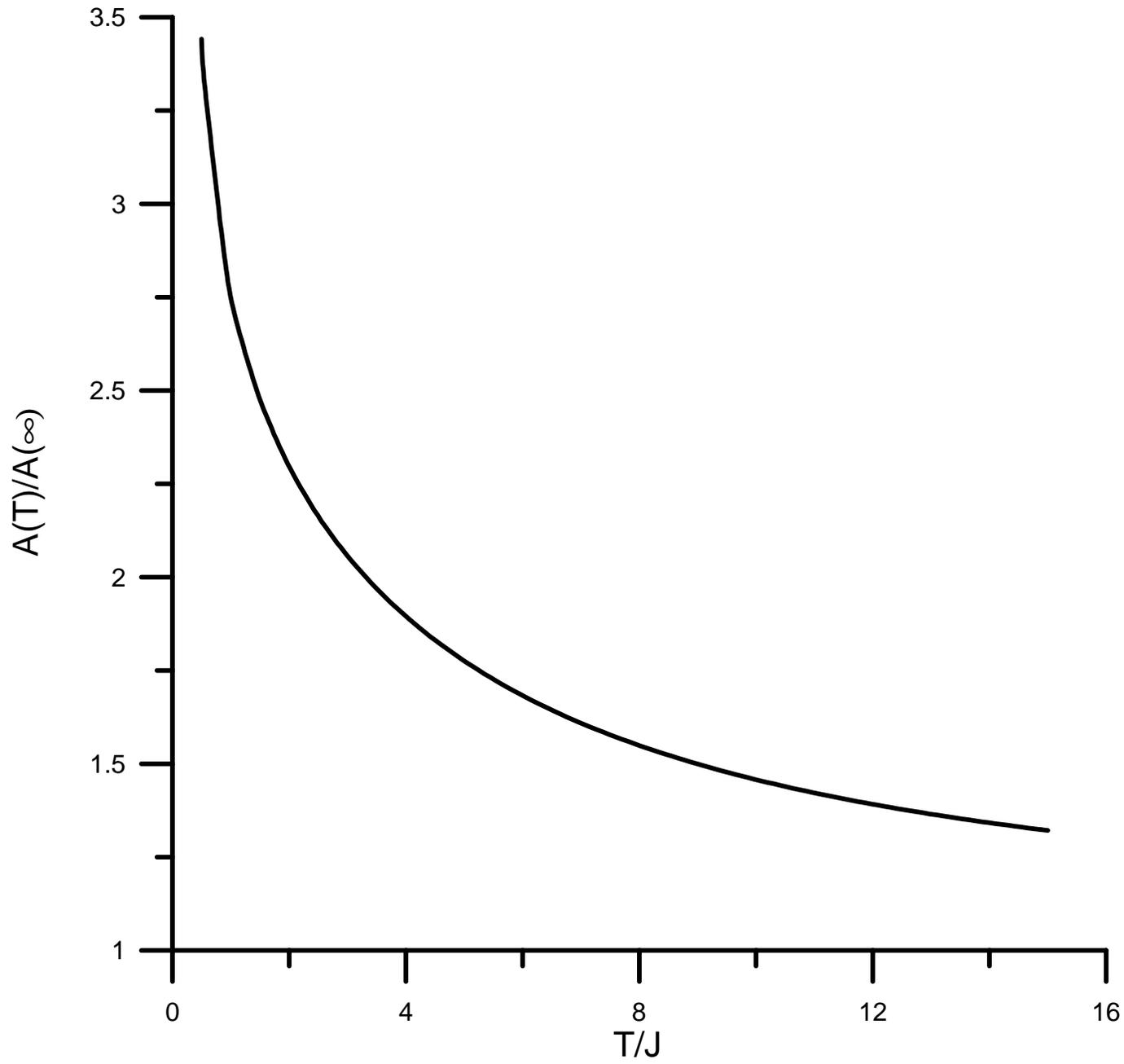

FIG 1. $A(T)/A(\infty)$ vs $k_B T/J$ for a nearest-neighbor Heisenberg antiferromagnet with $S = 3/2$ on a pyrochlore lattice. The peak in the susceptibility occurs for $T/J \approx 1.53$.

[1] S. T. Bramwell and M. J. P. Gingras, Science **294**, 1495 (2001).

[2] A. J. Garcia-Adeva and D. L. Huber, Physical Review B **63**, 174433 (2001).

[3] P. G. de Gennes, J. Phys. Chem. Solids **4**, 223 (1958).

[4] H. Mori, and K. Kawasaki, Progr. Theor. Phys. **27**, 529 (1962).

[5] H. Mori, Progr. Theor. Phys. **33**, 423 (1965).

[6] W. Marshall and S. W. Lovesey, *Theory of Thermal Neutron Scattering* (Oxford University Press, London, 1971), Chap. 13. Note that the exchange integrals in this reference differ from the integrals in our paper by a factor of $-2$.

[7] S.- H. Lee, C. Broholm, T. H. Kim, *et al*., Phys. Rev. Lett. **84**, 3718 (2000).

[8] S.- H. Lee, C. Broholm, W. Ratcliff, *et al.*, Nature **418**, 856 (2002).

[9] J. W. Lynn, Intl. Jour. Mod. Phys. B **12**, 3355 (1998).

[10] M. T. Causa, G. Alejandro, M. Tovar, *et al*., J. Appl. Phys. **85**, 5408 (1999).

[11] H. Martinho, N. O. Moreno, J. A. Sanjurjo, *et al*., Phys. Rev. B **64**, 024408 (2001).

[12] A. J. Garcia-Adeva and D. L. Huber, Physica B **320**, 18 (2002).